# Structure-Dependent Chemical Order Modification in Strained Alloy Nanoparticles

Yue Wang[1], Zibo Chen[1], Evropi Toulkeridou[2], Joseph Kioseoglou[3,4], and Panagiotis Grammatikopoulos[1,5,6*]

[1]*Materials Science and Engineering, Guangdong Technion – Israel Institute of Technology, Shantou, Guangdong 515063, China*

[2]*Universidad de Castilla-La Mancha, 13071 Ciudad Real, Spain*

[3]*School of Physics, Department of Condensed Matter and Materials Physics, Aristotle University of Thessaloniki, 54124 Thessaloniki, Greece*

[4]*Center for Interdisciplinary Research and Innovation, Aristotle University of Thessaloniki, 57001 Thessaloniki, Greece*

[5]*Guangdong Provincial Key Laboratory of Materials and Technologies for Energy Conversion, Guangdong Technion – Israel Institute of Technology, Shantou, Guangdong 515063, China*

[6]*Instituto Regional de Investigación Científica Aplicada (IRICA) and Departamento de Física, Universidad de Castilla-La Mancha, 13071, Ciudad Real, Spain*

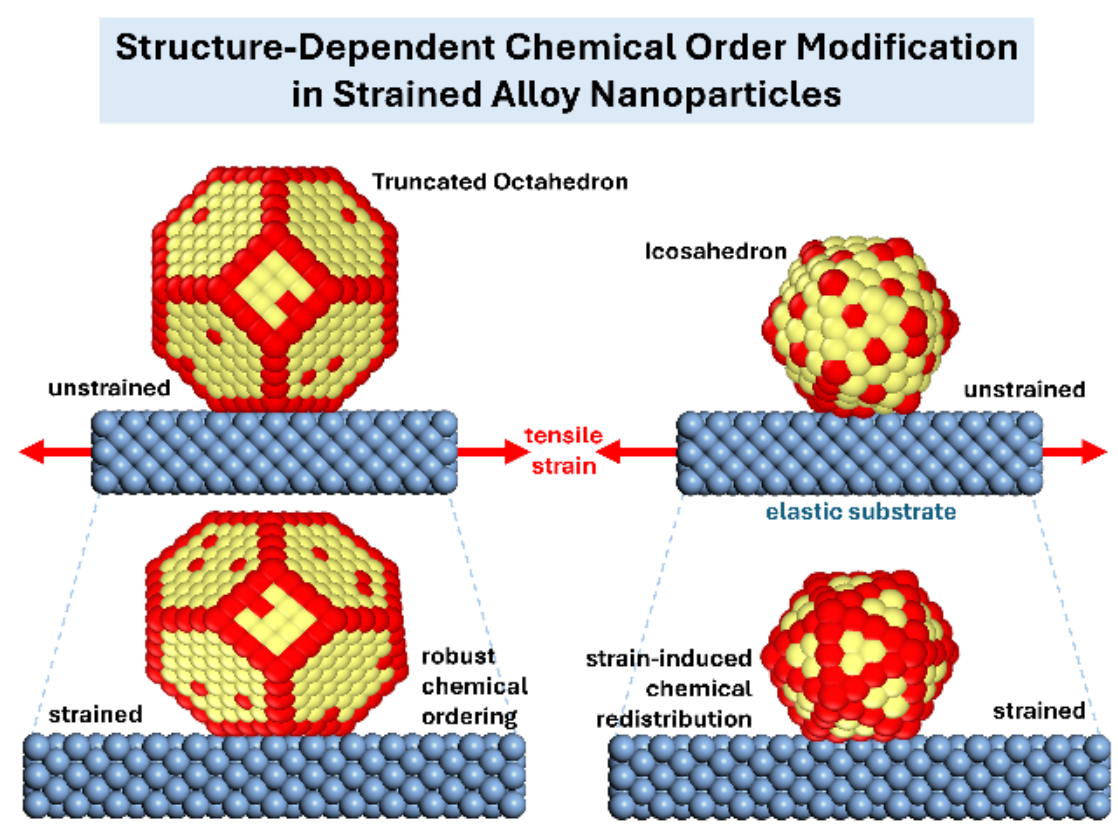

**ABSTRACT:** Alloy nanoparticles (nanoalloys) exhibit tuneable physicochemical properties that depend sensitively on their atomic arrangement, making control over chemical ordering a central challenge in nanomaterials design. While most theoretical studies consider nanoalloys in vacuum, practical systems are typically supported, where strong cluster–substrate interactions can introduce significant lattice strain. Here, we investigate strain as a control parameter for chemical ordering in bimetallic nanoalloys using atomistic molecular dynamics and Monte Carlo simulations. By imposing controlled tensile and compressive strain through an implicit anchored interface, we systematically probe the response of NiPt nanoparticles with distinct structural motifs. For truncated octahedral particles, we find that chemical ordering and segregation behaviour remain remarkably robust even under large strains, indicating that intrinsic thermodynamic preferences dominate. In contrast, icosahedral nanoparticles exhibit pronounced strain-induced chemical redistribution, with a significant increase in surface Ni concentration under tensile strain. This behaviour is attributed to the combined effects of intrinsic geometric frustration and a high fraction of undercoordinated sites in icosahedral structures. Our results demonstrate that strain can selectively modulate chemical ordering in nanoalloys in a structure-dependent manner, establishing a general framework for understanding strain-induced chemical ordering in nanoalloys.

## INTRODUCTION

Alloy nanoparticles (nanoalloys) have attracted considerable attention due to their tuneable physicochemical properties and wide-ranging applications in catalysis, sensing, optics, magnetism, etc. [1-7]. Their functional performance is governed not only by size and morphology but also by chemical ordering, i.e., the spatial arrangement of constituent elements within the particle [1,8]. In particular, the distribution of elements at the surface plays a decisive role in catalytic activity, making the ability to predict and control chemical ordering a central objective in nanoalloy research [9-11].

The equilibrium structure of nanoalloys is determined by a competition between several key factors, including enthalpy of mixing, surface energy differences, and strain energy arising from atomic size mismatch [12,13]. If all these factors work together, identifying this ordering is straightforward. However, they may compete with each other, leading to complex behaviour that is difficult to anticipate, especially since their influence changes with nanoparticle size [14-16]. As a result, the equilibrium structures of multi-elemental nanoparticles have been investigated thoroughly; however, to this day, this has mostly been done with the implicit assumption that the particles are in vacuum. While significant progress has been made in understanding nanoalloy structures, the vast majority of theoretical studies neglect the influence of the supporting environment [12].

In practical applications, on the other hand, nanoalloys are typically supported on substrates, where strong cluster–support interactions can induce partial or complete epitaxial alignment [17-19]. This interaction introduces lattice strain within the nanoparticle, even in cases of significant lattice mismatch, which can modify its structural and electronic properties [20,21]. While such effects have been extensively explored for monometallic nanoparticles [22], their impact on the chemical ordering of nanoalloys remains largely unexplored.

Most importantly, strain represents a potentially tuneable parameter capable of modifying the balance between competing energetic contributions in nanoalloys [23]. In particular, externally imposed strain through a deformable support may influence segregation behaviour, surface composition, and structural stability. However, whether strain can overcome intrinsic thermodynamic preferences and enable control over chemical ordering—and under what conditions—remains an open question.

Here, we address this question using atomistic simulations of a suitable nanoalloy system (namely, NiPt) subjected to controlled tensile and compressive strain via an implicit anchored interfacial layer. By comparing nanoparticles with distinct structural motifs, we demonstrate that the response to strain is strongly structure-dependent. While close-packed truncated octahedral (TO) nanoparticles exhibit robust chemical ordering under strain, icosahedral (Ih) nanoparticles undergo significant strain-induced compositional redistribution. These findings reveal a previously unrecognised interplay between geometric structure and strain, establishing new design principles for strain-engineered nanoalloys.

## RESULTS AND DISCUSSION

### Model System Selection and General Framework

The ability of strain to modify chemical ordering in nanoalloys depends sensitively on the balance between competing energetic contributions, including enthalpy of mixing, i.e., whether the atoms prefer to be surrounded by atoms of their own kind or different ones, surface energy differences between constituent elements, since demixing in nanoparticles is typically connected with surface segregation, and strain energy [13,24-26]. To probe this interplay, it is necessary to consider a system in which these contributions are of comparable magnitude, such that external perturbations can potentially shift the equilibrium chemical configuration.

In this work, we focus on NiPt nanoalloys as a representative model system. The Ni–Pt combination is particularly suitable for this purpose due to (i) its moderate negative enthalpy of mixing favouring alloy formation [27,28], (ii) significant surface energy differences between the constituent elements [29] promoting segregation, and (iii) moderate atomic size mismatch [30], introducing strain effects without overwhelming the other energetic contributions. Together, these characteristics place NiPt in a regime where chemical ordering is sensitive to perturbations, making it an ideal candidate for investigating strain-induced effects.

A detailed justification of the model system selection, including quantitative estimates of energetic contributions and comparison with other alloy systems, is provided in the **Supporting Information** (Section S1).

**Structural Response of Truncated Octahedral Nanoalloys under Strain**

To establish a baseline for the effect of strain on nanoalloy structure and chemical ordering, we first examine TO NiPt nanoparticles, which represent equilibrium morphologies compatible with the face-centred cubic (*fcc*) lattice. These structures are characterised by close-packed atomic arrangements and minimal intrinsic strain, making them ideal reference systems for assessing the influence of externally imposed deformation. Furthermore, they are commonly fabricated by gas-phase synthesis, as TO is the energetically favourable shape for *fcc* nanoparticles in the size regime investigated here [31,32]; larger nanoparticles would likely be less susceptible to strain engineering, anyway.

**Figures 1a** and **1b** show the structural evolution of TO nanoparticles of different sizes (containing 586 and 1289 atoms, respectively, roughly corresponding to 2.5 and 3.5 nm in diameter) under a range of imposed strains, spanning compressive to tensile regimes (from -30 to +50%). In all cases, the nanoparticles maintain their overall morphology, with only moderate distortions observed at higher strain magnitudes and mostly in the vicinity of the strain-imposing bottom layer (**Supporting Information**, Section S2). Notably, the atomic arrangement remains largely ordered in a manner consistent with equilibrium segregation tendencies (i.e., those at 0% strain).

A key observation is the persistence of Ni surface segregation across the entire strain range. Despite the significant lattice deformation imposed through the anchored interface, Ni atoms preferentially occupy surface sites, while the core remains enriched in Pt atoms (note that enrichment means atomic fraction higher than the nominal composition of the nanoparticle, not necessarily than that of the other constituent element). This behaviour reflects the dominant role of surface energy differences in determining chemical ordering, which is not overcome by externally applied strain within the explored range, exaggerated though it may be.

It should be noted that a number of theoretical studies indicate the opposite surface segregation tendency [1,33,34]. However, that tendency is in contrast to various experimental findings [35,36]. The demixing and surface occupation of the nanoalloy depends on the interatomic potential used for each study, and the underlying first-principles and experimental parameters on which it relies [37]. For consistency, in our system selection process elaborated in **Supporting Information** (Section S1) we selected values from as limited sources as possible. Accordingly, we opted for an interatomic potential that complies with these values and predicts Ni-enriched surfaces for the compositions investigated [38,39]. Nevertheless, the purpose of the present study is not to resolve the equilibrium segregation behaviour of NiPt, which remains sensitive to model assumptions, particle size, and thermodynamic conditions, but rather to investigate how externally imposed strain modifies the resulting chemical order. In that sense, the following discussion is fully self-consistent and can act as a guide for studies on other nanoalloy systems.

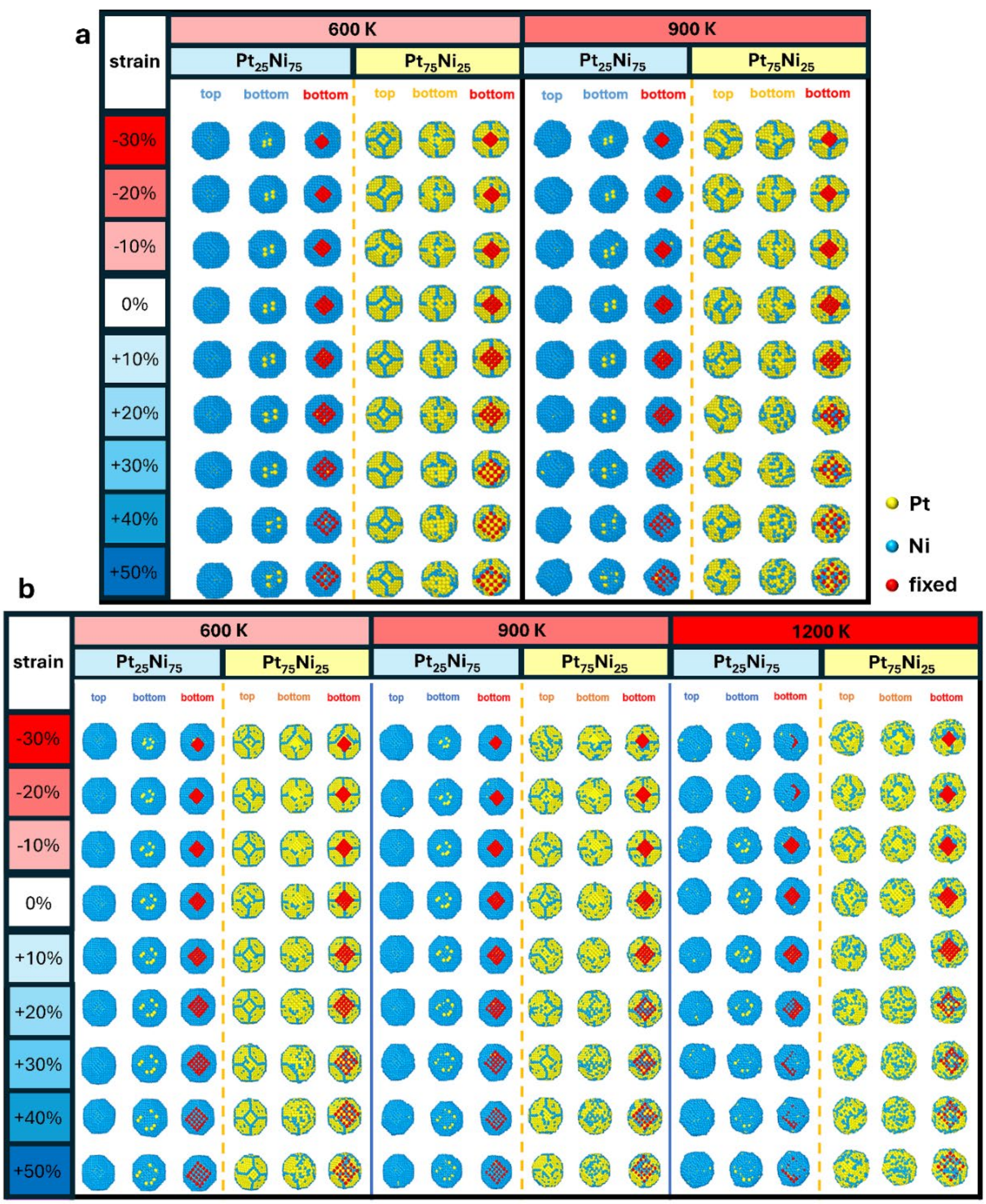

**Figure 1. Structural evolution of truncated octahedral NiPt nanoparticles under applied strain.** (a) Smaller nanoparticle (586 atoms) and (b) larger nanoparticle (1289 atoms) shown across compressive to tensile strain regimes. Ni and Pt atoms are distinguished by colour: blue for Ni and yellow for Pt. Each configuration at each temperature (600-1200 K) is viewed from two viewpoints, shown in corresponding columns: top view (away from the strained bottom layer) and bottom view (through the strained bottom layer). In the third column representing each configuration, bottom-layer frozen atoms are depicted red to stand out. Small and large nanoparticles are not to scale. The overall morphology and chemical distribution remain largely unchanged, indicating robust segregation behaviour despite the exaggerated strains imposed to the nanoparticles. Visualisations performed with OVITO [40].

## Quantitative Analysis of Chemical Ordering

To further assess the impact of strain on chemical ordering, we analyse radial concentration profiles and short-range order (SRO) parameters for TO nanoparticles as complementary measures of chemical distribution within the nanoparticles. The radial concentration profiles (Figure 2) show a consistent enrichment of Ni at the surface and Pt in the core across all strain values. Importantly, the shape and magnitude of these profiles remain essentially unchanged under both compressive and tensile strain, confirming the robustness of segregation behaviour. The SRO analysis (Figure 3) further supports this conclusion. The calculated Warren–Cowley parameters remain close to zero across the strain range, indicating a lack of chemical ordering beyond random mixing. No systematic trend with strain is observed, reinforcing the conclusion that strain does not induce ordering transitions in these systems.

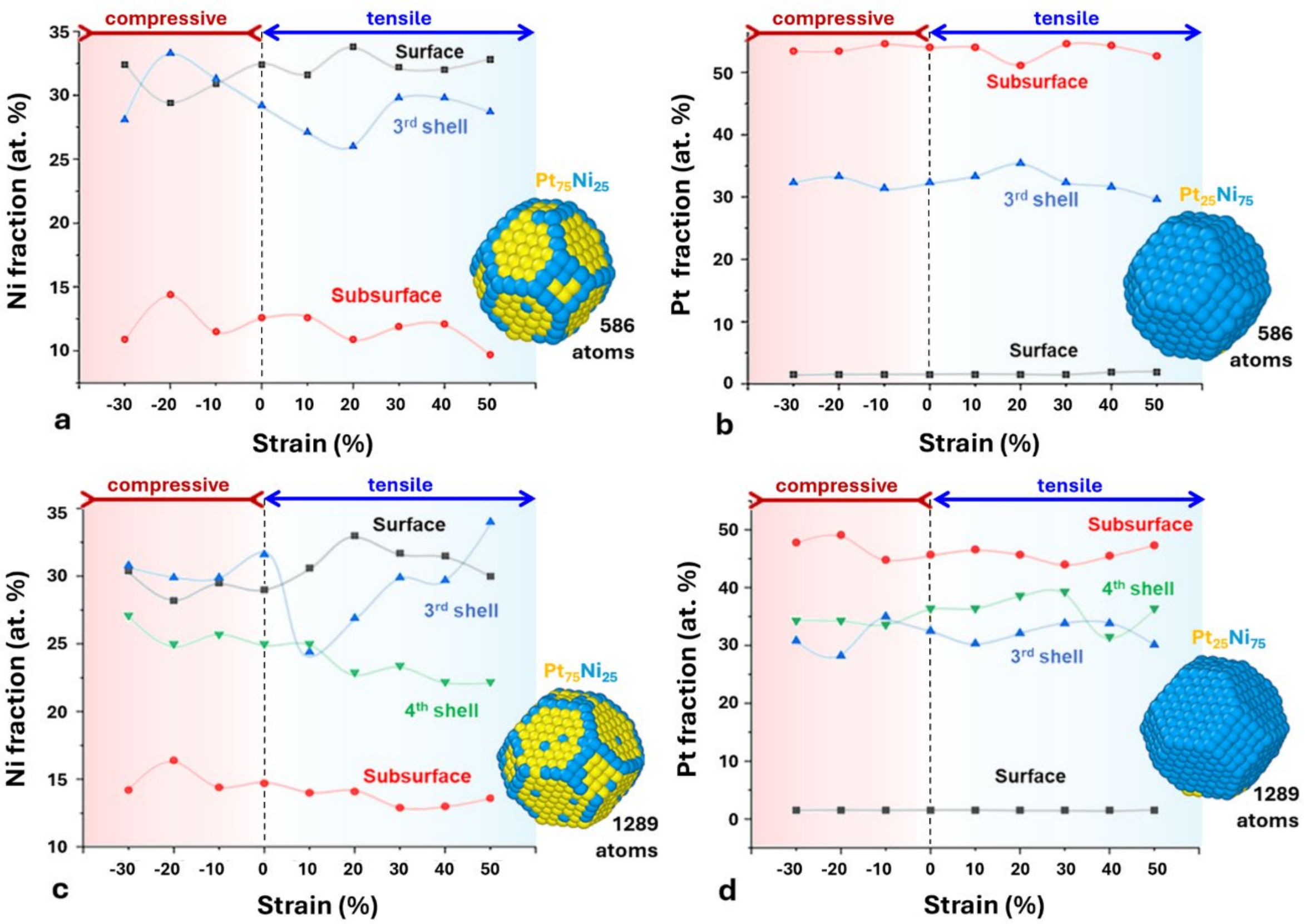

**Figure 2. Radial concentration profiles for truncated octahedral nanoparticles under different strain conditions.** (a-b) Pt-rich and Ni-rich configurations, respectively, of small nanoparticles (~2.5 nm in diameter). The radial concentration of the minority constituent atoms (Ni in the Pt-rich configuration and Pt in the Ni-rich configuration) is shown for the three outmost atomic layers. (c-d) Pt-rich and Ni-rich configurations, respectively, of large nanoparticles (~3.5 nm in diameter). The radial concentration of the minority constituent atoms is shown for the four outmost atomic layers. Lines are guides to the eye. All insets depict final configurations after relaxation at zero strain (not to scale). Pt and Ni atoms are depicted yellow and blue, respectively. Ni remains enriched at the surface, while Pt preferentially occupies the core. Profiles show negligible variation with strain. Visualisations performed with OVITO [40].

Together, these results demonstrate that, for structurally stable *fcc*-like nanoparticles, chemical ordering is largely insensitive to externally imposed strain. The close-packed geometry provides a deep energetic minimum, and the system responds to deformation primarily through elastic distortion rather than atomic rearrangement. This establishes a baseline in which strain alone is insufficient to alter chemical ordering, highlighting the importance of structural factors in enabling strain-induced effects.

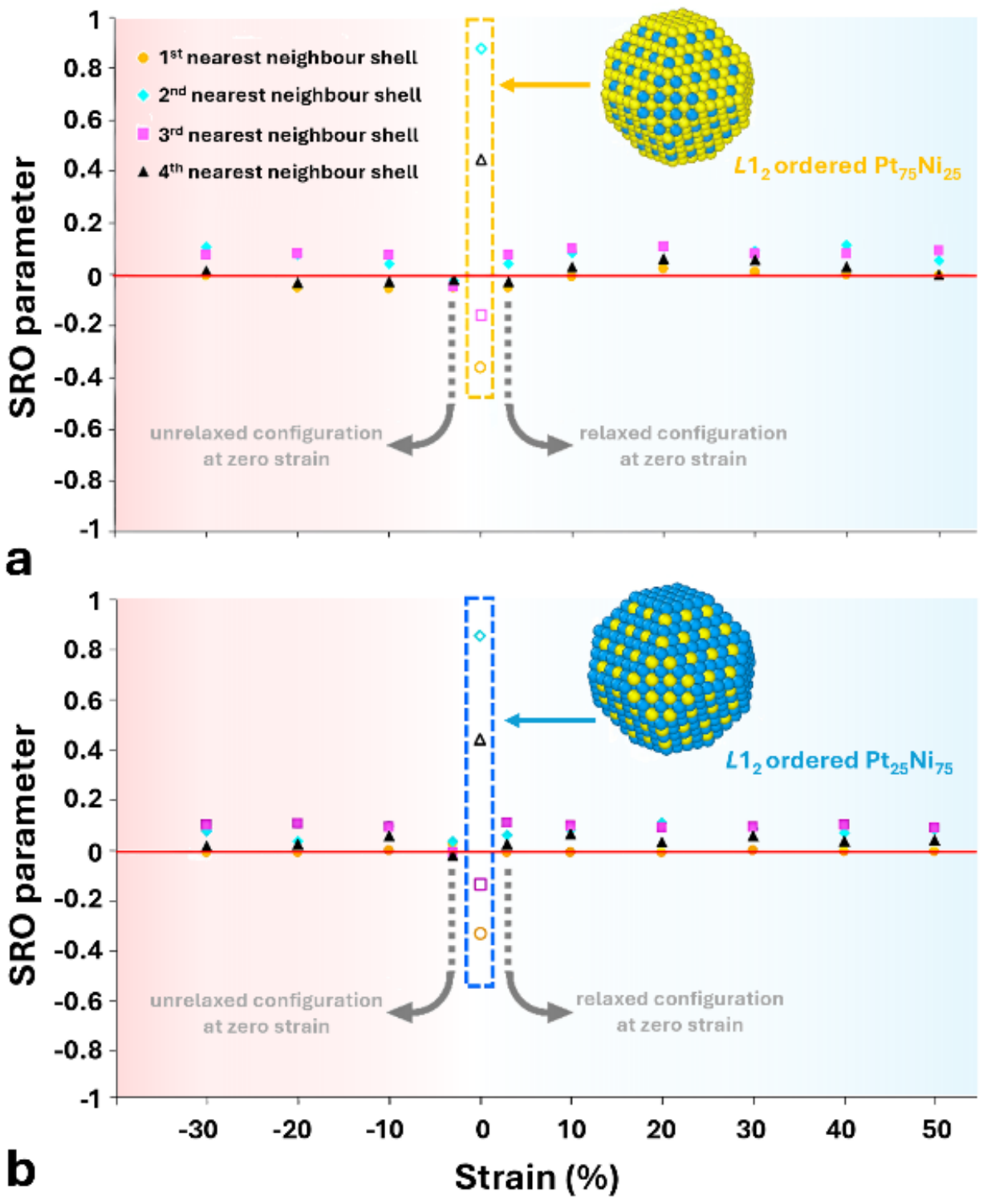

**Figure 3. Warren–Cowley SRO parameters calculated for four nearest-neighbour shells of the 3.5 nm (a) Pt-rich and (b) Ni-rich TO nanoparticles as a function of strain at 600 K.** As a reference, the insets show corresponding $L1_2$-ordered particles, the SRO parameters of which are indicated by open points inside the dotted rectangular frames. 2nd- and 4th-nearest neighbours have unambiguously positive SRO parameters, while 1st- and 3rd-nearest neighbours have negative SRO parameters, stemming from the crystallographic order of the clusters. In contrast, the SRO values of the initially random particles remain close to zero regardless of strain value, indicating random chemical mixing and absence of strain-induced ordering. Visualisations performed with OVITO [40].

## Strain-Induced Chemical Redistribution in Icosahedral Nanoalloys

In contrast to the robust behaviour observed for TO nanoparticles, Ih NiPt nanoalloys exhibit a markedly different response to externally imposed strain. Icosahedral structures are characterised by fivefold symmetry, which is incompatible with translational periodicity and results in intrinsic geometric strain. This inherent structural frustration, combined with a high fraction of undercoordinated atoms located at edges and vertices, creates an energetically perturbed configuration that is expected to be more sensitive to external perturbations.

Figure 4 presents the structural evolution of a small $Pt_{75}Ni_{25}$ Ih nanoparticle containing 309 atoms at 600 K under varying strain conditions. At such small sizes the Ih structure is more favourable than the TO one for systems following *fcc* crystallography [31,41]. Unlike the TO case, clear changes in chemical distribution are observed as a function of strain. In particular, tensile strain promotes the migration of Ni atoms toward surface sites (predominantly on edges, which form the vast majority of surface sites, anyway), resulting in a pronounced increase in surface Ni concentration. This effect is visually evident in the structural snapshots and is quantitatively confirmed by the corresponding compositional analysis.

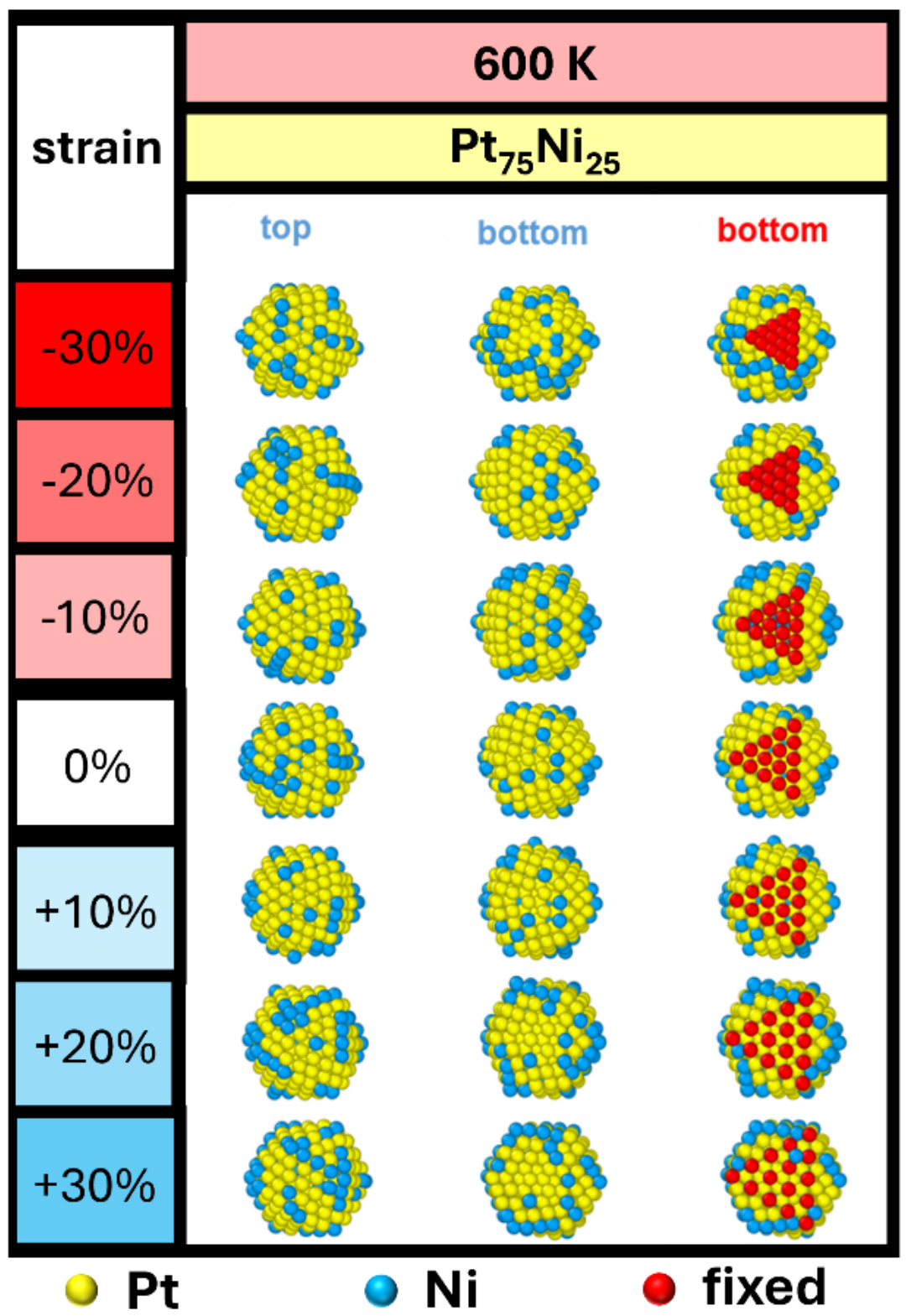

**Figure 4. Structural evolution of a small icosahedral $Pt_{75}Ni_{25}$ nanoparticle (309 atoms) under applied strain at 600 K.** Atomic configurations are shown across compressive to tensile strain regimes. Views are similar to those in Figure 1. Ni and Pt atoms are depicted blue and yellow, respectively. Tensile strain promotes redistribution of Ni atoms toward the nanoparticle surface, particularly at undercoordinated edge regions, whereas compressive strain largely suppresses this behaviour. Visualisations performed with OVITO [40].

The strain-induced redistribution observed in the structural snapshots is analysed quantitatively in Figure 5. Most importantly, Figure 5a demonstrates that tensile strain leads to a substantial increase in the fraction of Ni atoms occupying surface sites in the Ih nanoparticle. While Ni atoms already exhibit a moderate tendency for surface segregation in the relaxed zero-strain configuration (the surface concentration of Ni atoms increased from 25.9 at. % to 27.6 at. % after lattice relaxation), the application of tensile strain amplifies this behaviour considerably. At the highest applied tensile deformation, the surface Ni concentration increases markedly relative to the equilibrium configuration: as the bottom facet undergoes a 30% tensile strain, around 33 at. % of surface sites become occupied by Ni atoms, indicating that externally imposed strain actively modifies the equilibrium chemical distribution. In contrast, compressive strain largely suppresses this trend, producing only minor variations in surface composition.

Figures 5b and 5c further clarify how this redistribution occurs by distinguishing between undercoordinated edge sites and terrace-like surface sites. Under tensile strain, the fraction of Ni atoms occupying edge environments increases strongly, approaching nearly 50 at. % at the highest deformation. By contrast, the concentration of Ni atoms located at terrace surface sites exhibits a slight decrease with increasing tensile strain (within the statistical error margin). This distinction is significant because it demonstrates that Ni atoms do not simply segregate uniformly to the surface. Instead, tensile strain drives a redistribution of Ni atoms specifically toward undercoordinated surface environments, while occupation of more highly coordinated terrace sites becomes comparatively less favourable.

This behaviour provides direct insight into the mechanism underlying strain-induced chemical reorganisation in the Ih nanoparticles. Icosahedral structures possess substantial intrinsic geometric frustration due to their non-crystallographic fivefold symmetry, which requires expansion and distortion of the constituent facets relative to an ideal *fcc* arrangement. In essence, to construct an Ih motif, the twenty facets of the nanoparticle must be expanded to some extent. The application of additional tensile strain further intensifies this facet expansion. Under these conditions, segregation of the smaller Ni atoms toward the surface becomes energetically favourable; however, occupation of terrace-like surface environments would further destabilise the expanded facets. Therefore, terrace positions are filled by the larger Pt atoms. Accordingly, the segregating Ni atoms preferentially occupy edge and vertex regions, where the reduced coordination provides greater local structural flexibility and more efficient accommodation of the imposed strain.

The contrasting behaviour under compressive strain further supports this interpretation. Compression suppresses the tendency for Ni segregation and slightly reduces occupation of undercoordinated edge environments, consistent with relaxation toward a less expanded and therefore less frustrated structural state. More specifically, a nanoparticle under a 30% compression strain shows almost the same ratio of surface Ni atoms (~27.5 at. %) as its unstrained counterpart, whereas Ni edge occupation even decreases from 36.2 at. % (after relaxation) to 35.4 at. %. Taken together, these results demonstrate that the externally imposed strain does not merely perturb the nanoparticle elastically but instead couples directly to the intrinsically frustrated geometry of the Ih motif, enabling selective compositional redistribution among distinct classes of surface sites.

This mechanism contrasts sharply with the behaviour of TO nanoparticles, where the close-packed *fcc* geometry provides a deep and well-defined energetic minimum. In those systems, externally imposed strain is accommodated primarily through elastic distortion, while the equilibrium chemical ordering remains largely unchanged. The present results therefore demonstrate that strain-induced modification of chemical ordering is strongly structure-dependent and becomes particularly effective in geometrically frustrated nanoparticle motifs.

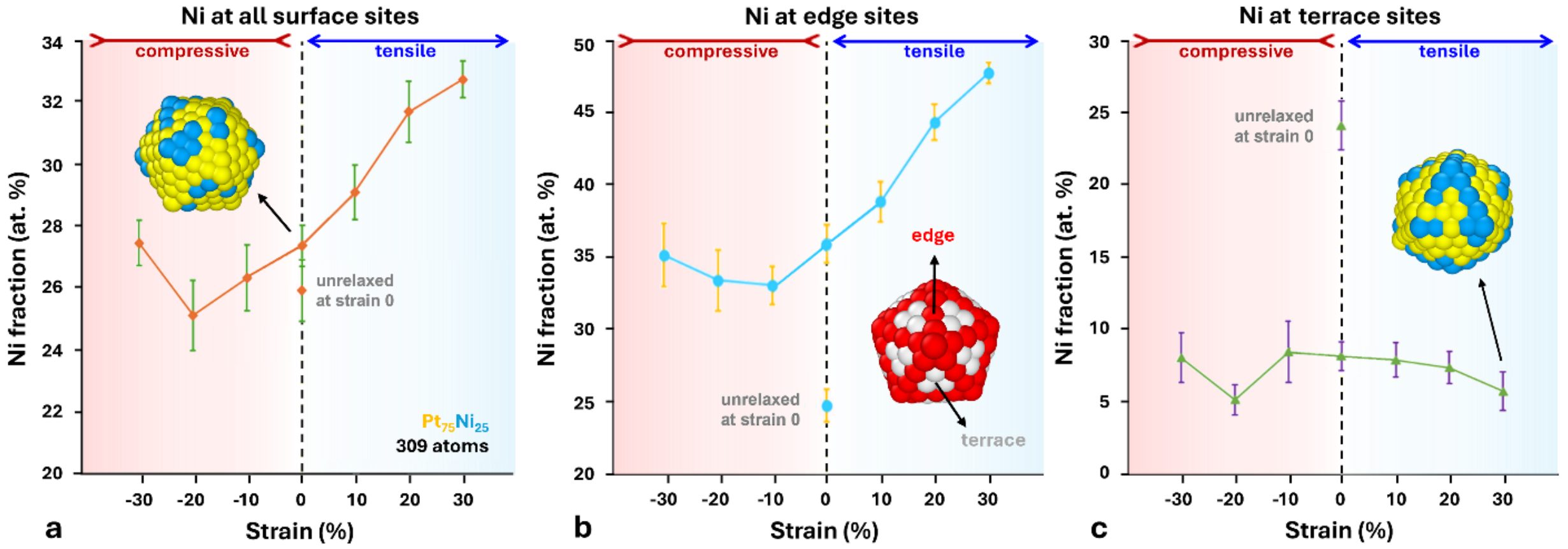

**Figure 5. Quantitative analysis of strain-induced chemical redistribution in an icosahedral $Pt_{75}Ni_{25}$ nanoparticle at 600 K.** (a) Fraction of Ni atoms occupying surface sites as a function of imposed strain, showing pronounced Ni surface enrichment under tensile deformation. (b) Fraction of Ni atoms occupying undercoordinated edge sites and (c) fraction of Ni atoms occupying terrace surface sites as a function of strain. Insets of panels (a-c) depict a relaxed unstrained nanoparticle, positions of edge and terrace sites, and a relaxed, 30% tensile-strained nanoparticle, respectively. Visualisations performed with OVITO [40]. Tensile strain promotes preferential redistribution of Ni atoms toward undercoordinated surface environments, while occupation of terrace-like surface sites decreases slightly, indicating selective strain accommodation within the intrinsically frustrated Ih structure.

### General Implications: Structure-Dependent Response to Strain

The combined results reveal a clear and general principle governing the response of nanoalloys to external strain. While strain is often considered a potential control parameter for modifying atomic arrangements, its effectiveness depends critically on the underlying structural motif of the nanoparticle. Although demonstrated here for NiPt, the underlying mechanism is expected to extend to alloy systems exhibiting comparable competition between mixing, segregation, and strain energies.

For close-packed, *fcc*-compatible structures such as truncated octahedra, chemical ordering is dominated by thermodynamic factors, and the system responds to strain primarily through elastic deformation. In contrast, geometrically frustrated structures such as icosahedra exhibit a fundamentally different behaviour. Their intrinsic strain and high density of undercoordinated sites create a shallow and highly perturbed energy landscape, enabling external strain to drive compositional rearrangement.

This distinction establishes a structure-dependent framework for strain engineering in nanoalloys. Rather than acting as a universal control parameter, strain selectively influences systems that are already energetically susceptible to perturbation. As a result, the combination of structural motif and external strain provides a route for tuning chemical ordering that is not accessible in equilibrium, unstrained systems.

## CONCLUSION

In this work, we have investigated the role of externally imposed strain on the structural and chemical ordering of NiPt nanoalloys using atomistic simulations. By introducing controlled tensile and compressive deformation through an anchored interface, we systematically probed the response of nanoparticles with distinct structural motifs.

For TO nanoparticles, which are structurally compatible with the *fcc* lattice, chemical ordering and segregation behaviour remain robust under strain. Radial concentration profiles and SRO analysis confirm that intrinsic thermodynamic driving forces dominate, and strain is accommodated primarily through elastic deformation without inducing compositional rearrangement.

In contrast, Ih nanoparticles exhibit pronounced strain-induced chemical redistribution. Tensile strain leads to a significant increase in Ni concentration at surface and edge sites, demonstrating that strain can actively modulate chemical ordering in these systems. This behaviour is attributed to the combined effects of intrinsic geometric frustration and the high fraction of undercoordinated atoms, which together create a shallow energy landscape that facilitates atomic rearrangement.

These results establish a clear structure-dependent response of nanoalloys to strain. While close-packed morphologies are resistant to strain-induced chemical modification, geometrically frustrated structures provide a viable pathway for tuning chemical ordering through mechanical means. This distinction highlights the importance of structural motif in determining the effectiveness of strain as a design parameter and provides a framework for the rational engineering of nanoalloy structures beyond equilibrium conditions.

More broadly, the present framework illustrates how macroscopically imposed strain can be translated into nanoscale control over atomic arrangement. By coupling external mechanical stimuli to internal chemical order, strain offers a pathway for directing nanoalloy structure beyond equilibrium constraints. This concept may provide guidance for future experimental efforts toward targeted synthesis of structurally tailored nanoparticles.

## METHODS

**Nanoparticle Models:** $Pt_{75}Ni_{25}$ and $Pt_{25}Ni_{75}$ truncated octahedral (TO) and icosahedral (Ih) alloy nanoparticles were considered, containing 586-1289 and 309 atoms, respectively. The initial structures were constructed with the open-source tool *NanoMaterialsCAD*, which facilitates the creation and manipulation of nano-objects based on input data containing the crystallographic description of the object [42]. Input data were retrieved from the *Materials Project* database [43].

**Atomistic Simulation Protocol:** Each system was equilibrated at elevated temperature (600-1200 K) to ensure sufficient atomic mobility, followed by combined MD/MC sampling to allow chemical redistribution. Atom swaps were implemented using a MC procedure based on the Metropolis algorithm, after every five MD steps in the canonical NVT ensemble. The timestep length was set to 1 femtosecond. Thermodynamic properties were sampled every $10^5$ timesteps. Each simulation ran for up to $10^7$ timesteps. Strain was imposed by constraining a subset of atoms representing the nanoparticle–support interface. Incremental controlled deformation of this layer introduced either tensile or compressive strain, mimicking epitaxial interaction with an implicit substrate. The strain range was selected to span both moderate and extreme deformation regimes (more discussion on model approximations in **Supporting Information**, Section S3). Each straining step (of 10% with respect to the relaxed lattice constant of the majority element) was followed by a consecutive relaxation phase, performed under constant temperature conditions. For statistics, 8 runs were performed for each Ih configuration. The final configuration after relaxation can be thought of as the result of a longer, pure MD simulation in the NVT ensemble, at the cost of losing information on the kinetic evolution of the system.

Interatomic interactions were described using an embedded atom method (EAM) potential parametrised for the Ni–Pt system [38,39], and validated for the reproduction of key bulk and surface properties, including lattice constants, cohesive energies, and segregation trends [44]. The simulations were carried out using the LAMMPS simulation package [45].

**Structural and Chemical Analysis:** Visualisation was performed using OVITO [40]. Chemical ordering was characterised using radial concentration profiles (computed by binning atoms according to their distance from the nanoparticle centre of mass and calculating local composition) and Warren–Cowley short-range order (SRO) parameters [46,47], defined for an $A_xB_{1-x}$ nanoalloy as:

$$\alpha_{AB}^{m} = 1 - \frac{n_m}{c_A C_m} = 1 - \frac{P_{AB}^{m}}{c_A} \qquad (1)$$

$n_m$ is the number of *A*-type atoms that are found in the *m*-th nearest-neighbour shell (around a reference *B*-type atom), of which the coordination number is $C_m$. $P_{AB}^{m}$ is the ratio between $n_m$ and $C_m$, which describes the average probability that the *m*-th nearest-neighbour site is occupied by an *A*-type atom. $c_A$ shows the average concentration of *A*-type atoms in the $A_xB_{1-x}$ system. By definition, a positive SRO implies clustering; a negative SRO implies that the *B* atoms tend to be as far away from each other as possible, i.e., forming an ordered inter-metallic compound; an SRO equal to zero marks the boundary of the miscibility gap associated with the existence of a random distribution of *B* atoms.

It should be pointed out that to define each neighbour shell around each atom in our strained nanoalloys, we use the number of atoms each shell should contain in an fcc lattice (namely, 12/6/12/8 nearest-neighbours for the first four shells, respectively) and not the actual atomic distances (which are distorted, anyway). To this end, we used an open-source Python script to compute the Warren-Cowley SRO parameters of our nanoalloys [48,49]. More details on this treatment can be found in **Supporting Information** (section S4).

## ASSOCIATED CONTENT

### Supporting Information

The Supporting Information is available free of charge at DOI:

Supporting Information Document (containing Figures S1 to S7, Tables S1 to S6)

## AUTHOR INFORMATION

### Corresponding Author

*Email: p.grammatikopoulos@uclm.es

### Notes

The authors declare no competing financial interests.

## ACKNOWLEDGMENTS

PG was supported by a Beatriz Galindo Senior Fellowship (BG23/00144). Work performed at GTIIT was supported by funding from the Guangdong Technion Israel Institute of Technology and Guangdong Provincial Key Laboratory of Materials and Technologies for Energy Conversion, MATEC (No. MATEC2022KF001). The authors are grateful to Profs. G. Mpourmpakis and P. K. Chrysanthis (University of Pittsburgh) for providing access to the MetalNanoDB database, which served as a first tool for our materials screening.

# Supporting Information

### S1. Justification of Model System Selection: Suitability of the Ni–Pt System

The selection of a suitable model nanoalloy system is critical for investigating the potential of strain to influence chemical ordering. For strain-induced effects to be observable, the energetic contributions governing chemical order must be of comparable magnitude, such that external perturbations can alter their relative balance.

Three primary factors determine chemical ordering in nanoalloys:

1. **Enthalpy of Mixing ($\Delta H_{mix}$)**

   This term reflects the preference for homoatomic versus heteroatomic bonding. Systems with strongly negative $\Delta H_{mix}$ tend to form ordered alloys, while systems with strongly positive $\Delta H_{mix}$ favour phase separation.

2. **Surface Energy Differences ($\Delta\gamma$)**

   In nanoparticles, surface segregation is driven by differences in surface energy between constituent elements, with the lower surface energy element preferentially occupying surface sites (Tables S1-S2).

3. **Strain Energy ($\Delta E_{strain}$)**

   Strain arises from atomic size mismatch and structural constraints. In bulk systems, this contribution is often secondary, but it can become significant in nanoscale systems or under external perturbation.

**Table S1. Surface energies of metals with *fcc* structures.** For comparison, both truncated octahedra and icosahedra are considered. The data for surface energies of {111} and {100} facets is according to reference [S1].The total surface energy of truncated octahedra was calculated as the sum of the surface energies of eight {111} facets and six {100} facets divided by the total surface area, whereas for icosahedra the same quantity was obtained by dividing the sum of the surface energies of twenty {111} facets by the total surface area of icosahedra.

| Metal | Surface energy ($J/m^2$) | | | |
|---|---|---|---|---|
| | Single (1 1 1) facet | Single (1 0 0) facet | Truncated octahedron | Icosahedron |
| Ag | 1.172 | 1.2 | 1.178 | 1.172 |
| Au | 1.283 | 1.627 | 1.360 | 1.283 |
| Pd | 1.92 | 2.326 | 2.011 | 1.92 |
| Cu | 1.952 | 2.166 | 2.000 | 1.952 |
| Ni | 2.011 | 2.426 | 2.104 | 2.011 |
| Pt | 2.299 | 2.734 | 2.396 | 2.299 |
| Rh | 2.472 | 2.799 | 2.545 | 2.472 |
| Ir | 2.971 | 3.722 | 3.139 | 2.971 |
| Mn | 3.1 | | | 3.1 |

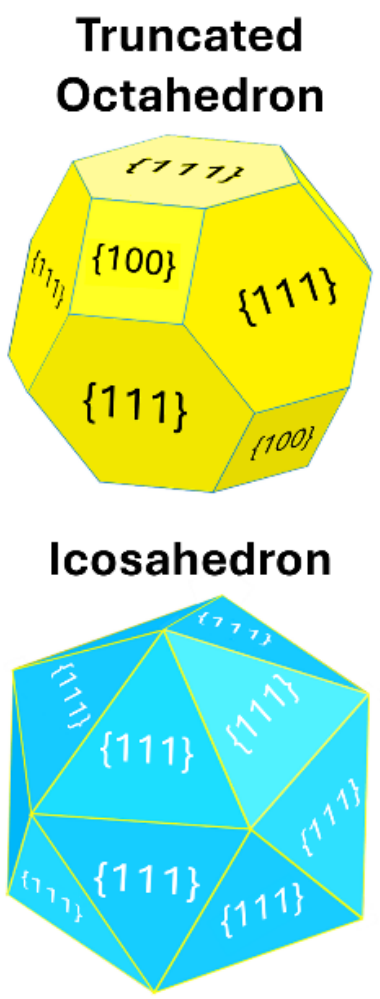

**Table S2. Surface energies of metals with *bcc* structures.** For comparison, both truncated octahedra and icosahedra are considered. The data for surface energies of {111} and {100} facets is according to reference [S1].

| Metal | Surface energy ($J/m^2$) | | | |
|---|---|---|---|---|
| | Single (1 1 1) facet | Single (1 0 0) facet | Truncated octahedron | Icosahedron |
| Fe | 2.43 | 2.222 | 2.383 | 2.43 |
| Nb | 2.685 | 2.858 | 2.724 | 2.685 |
| Ta | 3.084 | 3.097 | 3.087 | 3.084 |
| V | 3.258 | 3.028 | 3.206 | 3.258 |
| Cr | 3.505 | 3.979 | 3.611 | 3.505 |
| Mo | 3.454 | 3.837 | 3.540 | 3.454 |
| W | 4.005 | 4.635 | 4.146 | 4.005 |

Typically, to predict whether two elements can dissolve into each other forming a solid solution, the Hume-Rothery criteria are employed [S2]. These empirical rules dictate that high solid solubility requires that the atomic radii of the two elements must differ by less than 15%, and both elements must share the same crystal structure; further, their electronegativities and valencies must be similar. These rules serve as a fundamental screening tool for materials scientists in alloy design. If a mixture fails one or more of these criteria, the resulting lattice strain typically limits solubility. A list of values related to the Hume-Rothery rules of various materials is shown in Table S3.

**Table S3. Atomic radii, electronegativity values, and valencies for metals that are selected as candidates.** The values of atomic radii and valencies are taken from reference [S3]; the values of electronegativity are taken from reference [S4].

| Candidate metal | Structure | Atomic radius (Å) | Electronegativity (Pauling unit) | Valency |
|---|---|---|---|---|
| Cu | fcc | 1.296 | 1.4 | 5.44 |
| Pd | fcc | 1.373 | 2.2 | 6 |
| Pt | fcc | 1.385 | 2.3 | 6 |
| Ni | fcc | 1.244 | 1.9 | 6 |
| Rh | fcc | 1.342 | 2.3 | 6 |
| Ir | fcc | 1.355 | 2.2 | 6 |
| Mn | fcc | 1.306 | 1.6 | 4.6 |
| Fe | bcc | 1.260 | 1.8 | 5.78 |
| Nb | bcc | 1.456 | 1.5 | 5 |
| Ta | bcc | 1.457 | 1.4 | 5 |
| V | bcc | 1.338 | 1.6 | 5 |
| Mo | bcc | 1.386 | 2.2 | 6 |
| Cr | bcc | 1.357 | 1.7 | 3.6 |
| W | bcc | 1.394 | 2.3 | 6 |

With a prerequisite of weak mixing tendency, the interplay between surface and strain energies may play a critical role that dictates the formation of a core-shell configuration. Importantly, these two factors might work either cooperatively or competitively, depending on alloy composition. When surface energy and atomic size act cooperatively, they typically result in core-shell bimetallic nanoparticles [S5]. Therefore, a metal with both a higher surface energy and a smaller atomic size favourably occupies the core region, where strong binding among the constituent atoms and the relief from compressive strain occur simultaneously. In contrast, the surface energy and atomic size produce opposite effects when the smaller metal atoms also happen to possess lower surface energy; this is referred to as the competitive condition. A relative ordering of a number of transition metals with respect to these two critical factors is shown in Table S4. The decision to rank them in ascending and descending orders for surface energy and atomic size, respectively, was made because it indicates the collaborative or competing nature of the two criteria. When a metal occupies the same (or nearby) horizontal lines in this table for the two criteria (e.g., Ag), this implies a probable collaborative effect; when a metal ranks high in the one criterion but low in the other (e.g., Ni), it is a potential candidate for our model system.

**Table S4. Relative ranking of transition metals with respect to their potential surface segregation tendencies based on surface energies and atomic sizes.** Left columns: metals with *fcc* structures. Right columns: metals with *bcc* structures.

| *fcc* transition metals | | | | *bcc* transition metals | | | |
|---|---|---|---|---|---|---|---|
| Property | | Increasing Surface Energy | Decreasing Atomic Size | Property | | Increasing Surface Energy | Decreasing Atomic Size |
| Rank according to likelihood of surface segregation | 1 | Ag | Ag | Rank according to likelihood of surface segregation | 1 | Fe | Nb, Ta |
| | 2 | Au | Au | | 2 | Nb | W |
| | 3 | Pd | Pt | | 3 | Ta | Mo |
| | 4 | Cu | Pd | | 4 | V | V |
| | 5 | Ni | Ir | | 5 | Cr | Cr |
| | 6 | Pt | Rh | | 6 | Mo | Fe |
| | 7 | Rh | Cu | | 7 | W | |
| | 8 | Ir | Mn | | | | |
| | 9 | Mn | Ni | | | | |

Therefore, in selecting metals for bimetallic nanoparticles with adjustable chemical order, the combination principle is: small atoms with low surface energies and large atoms with high surface energies. According to this scheme, in Table S5 we propose a series of combinations that could lend themselves for potential chemical ordering rearrangement.

**Table S5. Candidates of bimetallic combinations subject to the competitive criterion (in red).** (a) *fcc* structure; (b) *bcc* structure.

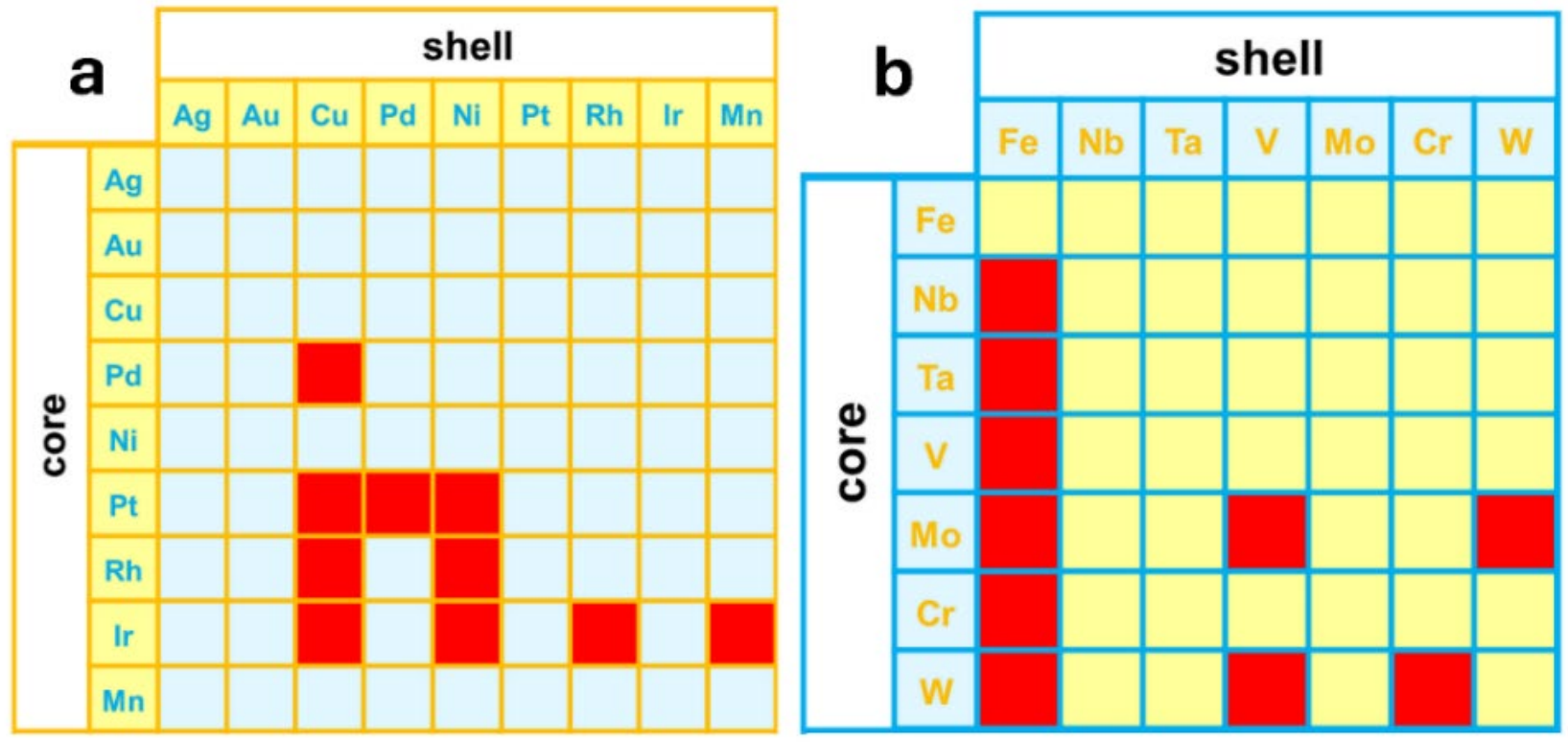

We should emphasise that the bimetallic combinations in Table S5 are proposed based on the competitive interplay between their surface energy and atomic size, whereas their fulfilment of the Hume-Rothery conditions is a necessary prerequisite (for simplicity we only consider combinations of metals of the same packing type). The difference in atomic size is considered a priority since it is the one most related to strain[1]. As Table S6 shows, except for the combinations of Nb-Fe and Ta-Fe, the difference in atomic size is less than 15% for most candidates. This suggests a significant potential for forming a well-mixed nanoalloy in these bimetallic candidates.

**Table S6. Relative differences in atomic radii for candidate metals from Table S5.** *fcc* structures (left); *bcc* structures (right).

| Candidate combination | Difference in atomic radii (%) |
|---|---|
| Pd-Cu | 7.6 |
| Pt-Cu | 8.5 |
| Rh-Cu | 5.2 |
| Ir-Cu | 5.8 |
| Pd-Pt | 0.9 |
| Pt-Ni | 11.3 |
| Rh-Ni | 7.9 |
| Ir-Ni | 8.9 |
| Ir-Rh | 1.0 |
| Ir-Mn | 3.7 |

| Candidate combination | Difference in atomic radii (%) |
|---|---|
| Nb-Fe | 15.5 |
| Ta-Fe | 15.6 |
| V-Fe | 6.2 |
| Mo-Fe | 10 |
| Cr-Fe | 7.7 |
| W-Fe | 10.7 |
| Mo-V | 3.6 |
| W-V | 4.2 |
| W-Mo | 0.6 |
| Nb-Ta | 0.1 |

The above investigation of the suitability of those bimetallic candidates is qualitative. There also exist a number of studies that computed and predicted the mixing-demixing trends in bimetallic nanoparticles, presenting a clearer viewpoint for potential core-shell preferences, e.g., references [S6] and [S8]. From these studies, combined with our bimetallic candidates of Table S5, four bimetallic systems were initially shortlisted: namely, Ni-Pt, Pt-Cu, Pt-Pd (*fcc*), and Mo-Fe (*bcc*). From these systems, Pt-Pd was subsequently excluded because, in contrast to Ni-Pt, Pt-Cu, and Mo-Fe, where the atomic sizes of the two metal species have a relatively large difference (11.3%, 8.5%, and 10%, respectively), thus resulting in a non-trivial probability of lattice mismatch, Pt-Pd presents a negligible difference in atomic size (only 0.9%), in spite of the Pt atom appearing larger in Table S4.

[1] On the other hand, the effect of electronegativity is controversial. The Hume−Rothery rules state that similar electronegativity is favourable in forming alloys, while others argue that a significant difference in electronegativity may also facilitate good mixing due to charge transfer from less to more electronegative metals [S6]. Eom *et al.* predicted the core-shell preference for selected bimetallic nanoparticles and their simulation results supported the latter argument [S7].

The Ni–Pt alloy system was finally selected because it satisfies the criteria for balanced energetic competition:

**Enthalpy of Mixing:** The enthalpy of mixing for Ni–Pt is moderately negative across a broad composition range, indicating a weak tendency toward alloy formation (at low temperatures) or solid solution (at high temperatures) rather than strong ordering or phase separation. This ensures that chemical ordering is not rigidly fixed by bulk thermodynamics.

**Surface Energy Contrast:** Ni exhibits a lower surface energy than Pt, leading to a natural tendency for Ni surface segregation in equilibrium configurations. However, this tendency competes with other effects at the nanoscale, particularly in small particles where coordination effects are significant.

**Atomic Size Mismatch:** Ni and Pt exhibit a moderate but significant atomic size mismatch (~10%), which introduces strain effects without dominating the energetics. This makes the system sensitive to additional externally imposed strain.

To contextualise the choice of NiPt, it is useful to compare with other classes of alloys:

- **Strongly ordering systems (e.g., Pt–Ti)**

  Large negative $\Delta H_{\mathrm{mix}}$ leads to robust ordering, which is unlikely to be modified by strain.

- **Strongly segregating systems (e.g., Ag–Ni)**

  Large positive $\Delta H_{\mathrm{mix}}$ and surface energy differences lead to phase separation, making strain effects secondary.

- **Nearly ideal systems (e.g., Ni–Pt)**

  Weak $\Delta H_{\mathrm{mix}}$ allows competition between energetic terms, making them ideal for probing external control mechanisms.

---

### S2. Structural Distortions and Distortion Propagation in Strained Nanoparticles

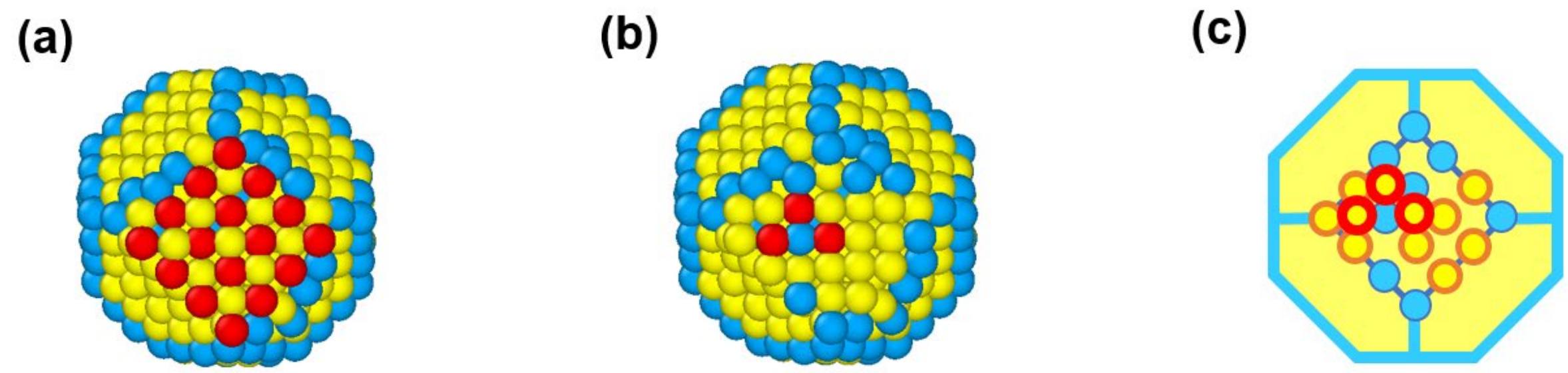

**Figure S1.** **Atomic sinking and leaching viewed through the bottom layer of a relaxed nanoalloy particle**. (a) Exemplar snapshot of 586-atom $Pt_{75}Ni_{25}$ particle (600 K, at 30% tensile strain), with fixed bottom atoms marked in red. (b) Same snapshot as (a), but with leaching atoms in red. (c) Simplified schematic illustration of atomic sinking and leaching; three leaching atoms are drawn with red outline. Atomic sinking is based on the Poisson effect (see Figure S2), but atomic leaching is an artifact of the simulation setup, and would not occur in the presence of a real cluster-support interface.

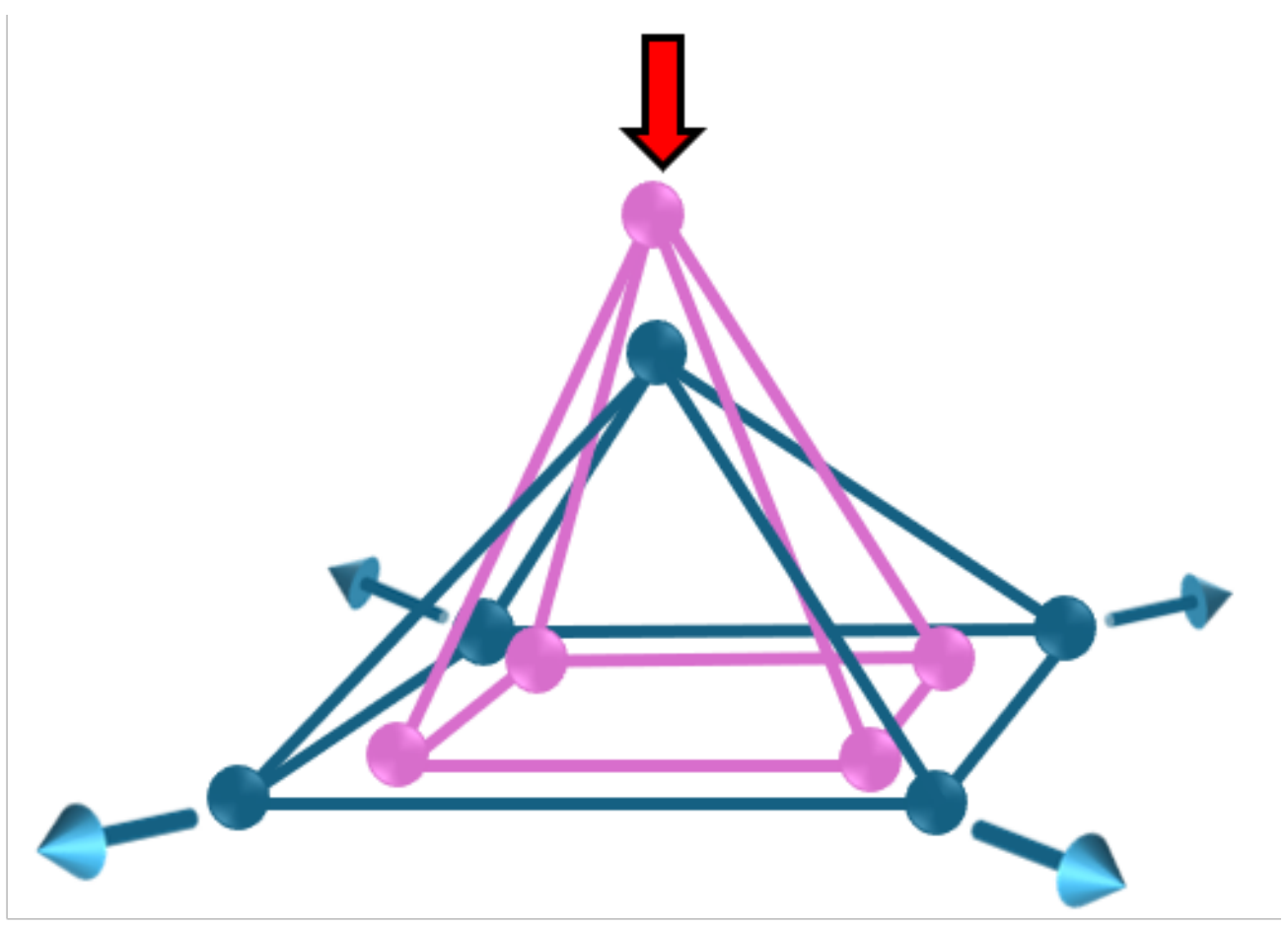

**Figure S2. Schematic representation of the out-of-plane atomic relaxation induced by biaxial tensile strain applied to the bottom layer, as expected from the Poisson effect.** As a result, the aforementioned atomic sinking and leaching ensues. Pink spheres and bonds correspond to the unstretched state; blue spheres and bonds correspond to the stretched state.

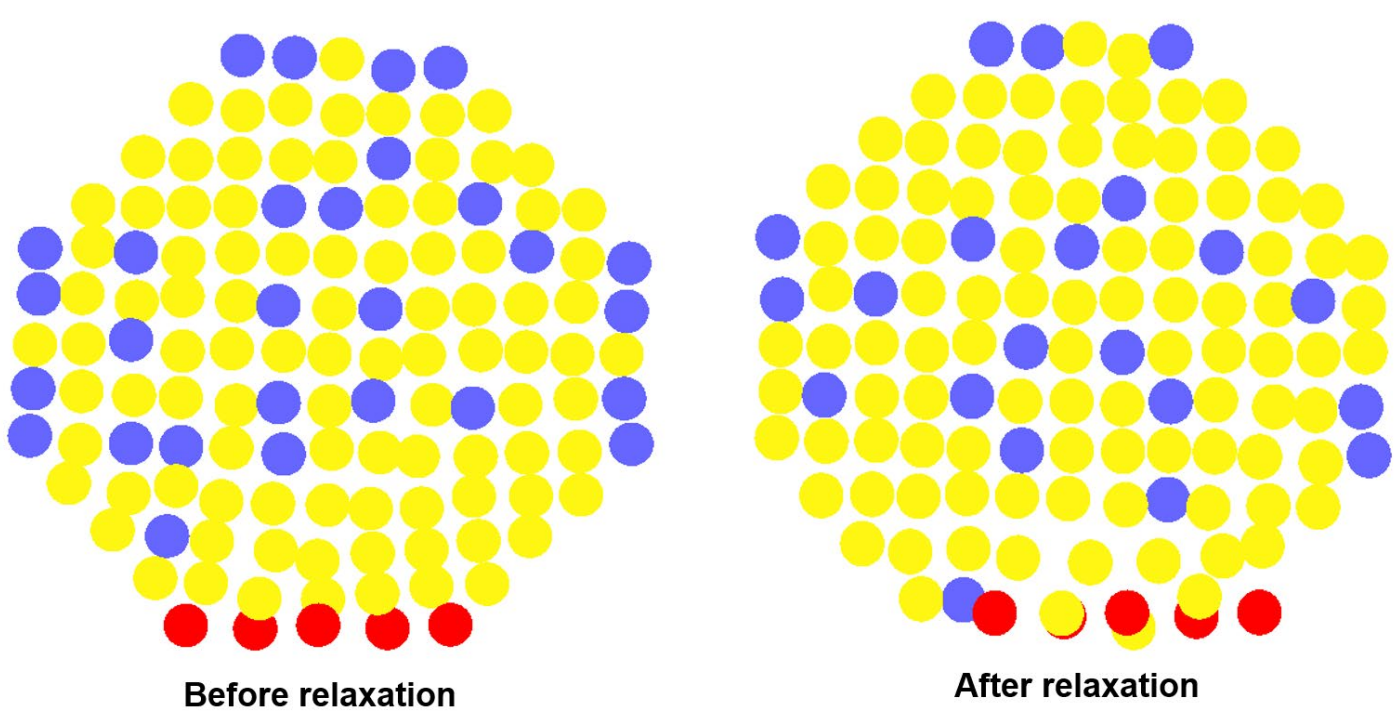

**Figure S3. Simplified illustrated comparison of structure before and after relaxation.** Note the second-from-bottom layer atoms moving downwards to approach the bottom level and partly conceal the view to the fixed bottom atoms (in red).

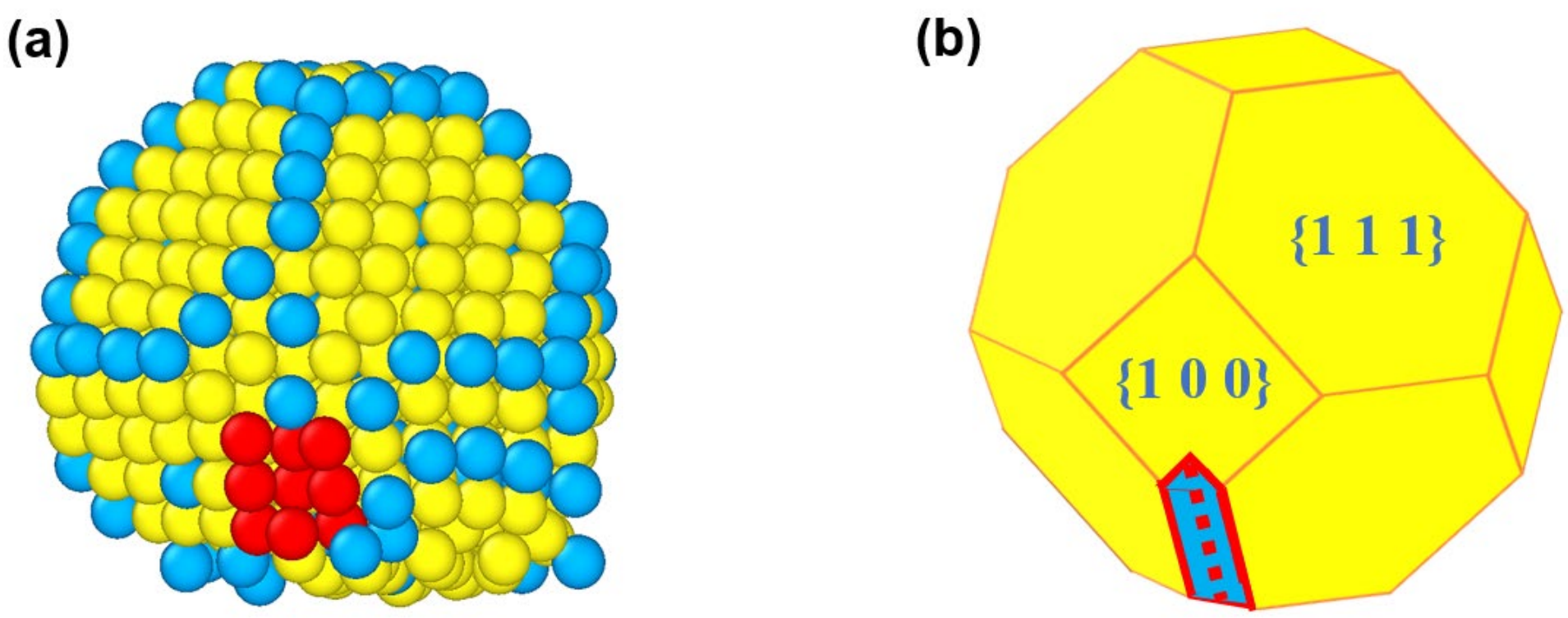

**Figure S4. Appearance of new {110} facet.** (a) Exemplar snapshot of relaxed 586-atom $Pt_{75}Ni_{25}$ particle (at 600 K and 30% tensile strain). {110}-facet atoms are indicated in red. (b) Schematic illustration highlighting the emerging {110}-oriented facet.

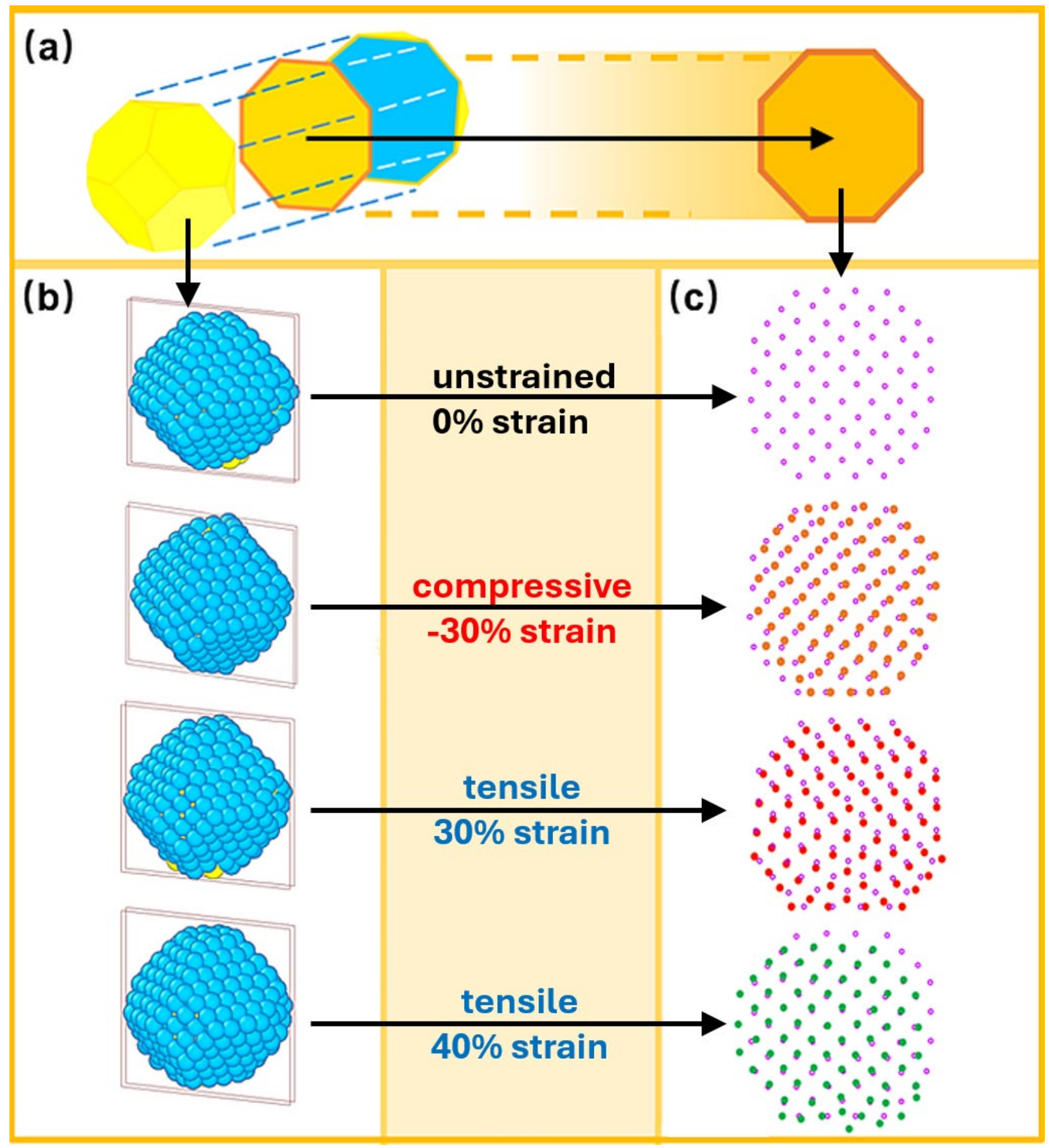

**Figure S5. Inspection of strain propagation in anchored nanoalloys**. (a) Schematic illustration of slicing a truncated octahedral particle. (b) Snapshots of relaxed nanoalloys, an equatorial one-atomic-layer thick slice of which is shown in (c). Atomic positions are depicted as purple hollow circles, orange solid circles, red solid circles, and green solid circles in nanoparticles under 0%, -30%, 30%, and 40% strain, respectively.

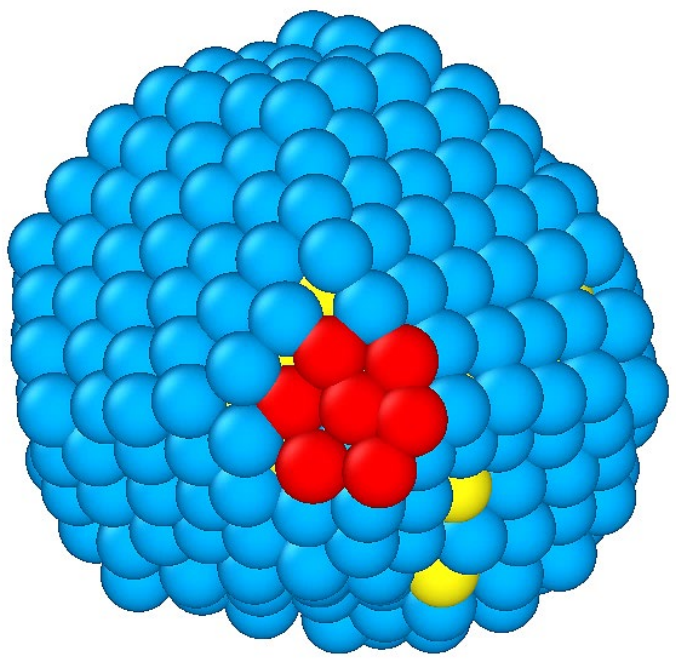

**Figure S6. Hexagon-shaped atomic arrangement of a distorted {100} facet.** Exemplar snapshot of 586-atom $Pt_{25}Ni_{75}$ nanoalloy particle (at 900 K, and 40% tensile strain). The atoms reconfiguring to hexagonal arrangement are highlighted in red.

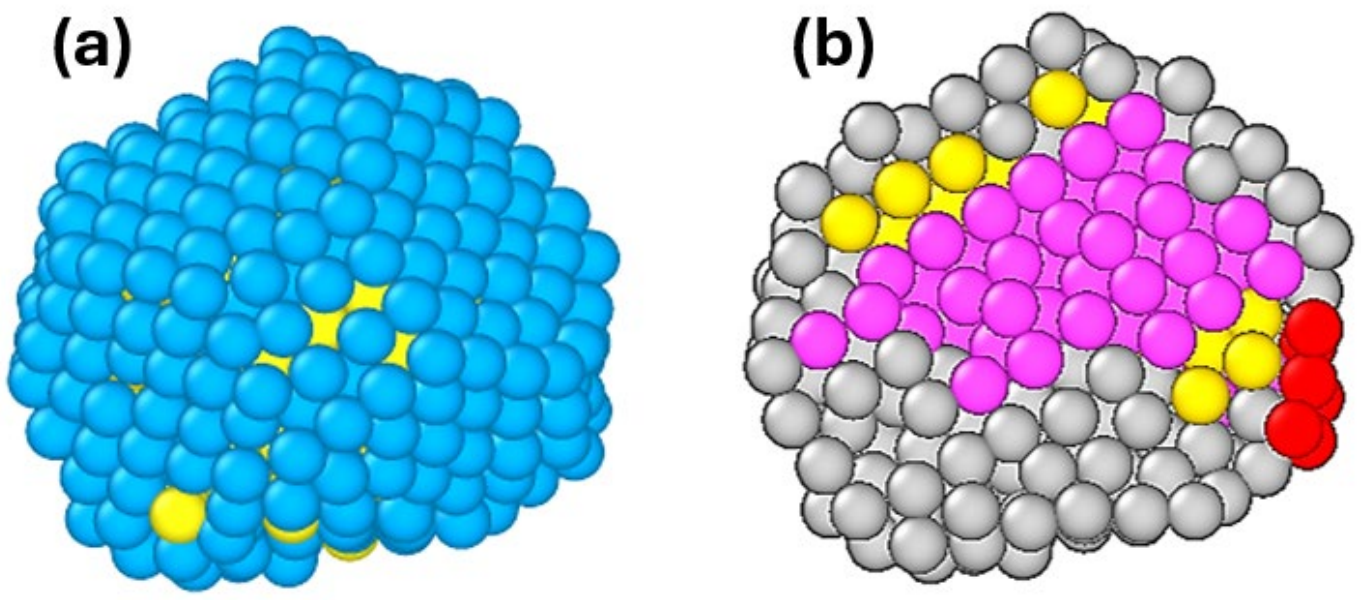

**Figure S7. Distortion inside a strained nanoalloy particle examined via common neighbour analysis (CNA).** (a) Nanoparticle snapshot. (b) Sliced view of CNA result, where atoms in the distorted stacking, *fcc*-stacking, and unidentified structural sites are depicted yellow, purple, and grey, respectively.

---

## S3. Fundamental Model Approximations

Two aspects of the model framework warrant clarification: (i) the implicit treatment of the support and (ii) the strain regime used to accommodate lattice mismatch between the nanoalloy and substrate.

First, the model assumes an "anchored" nanoparticle without explicitly including the substrate. This abstraction isolates strain-induced structural transformations by eliminating extrinsic complexities. However, two substrate-mediated effects are neglected. (1) Variations in surface/interface energy: analogous to adsorbate effects, substrate atoms can modify nanoparticle surface energetics and induce compositional segregation toward the interface. (2) Electronic interactions: charge transfer from the substrate may significantly alter nanoparticle morphology.

Second, the applied strain range is intentionally exaggerated to probe extreme structural responses beyond realistic physical limits. The model presumes coherent epitaxial alignment of the nanoparticle's bottom layer with the substrate, maintained under large imposed strains. In practice, such coherence is limited by lattice relaxation and the formation of misfit dislocations. While small mismatches (up to 7%) may be accommodated elastically below a critical thickness, larger strains promote dislocation formation, reducing interfacial coherence and ultimately leading to detachment [S9]. Consequently, sustaining large epitaxial strains—particularly at the highest values considered here—is unlikely for most materials, as they exceed typical elastic limits and introduce significant deviations from idealised behaviour.

Overall, the present study should be regarded as a proof-of-concept. Despite its simplified framework, it provides mechanistic insight into strain-driven structural evolution in alloy nanoparticles.

---

## S4. Technicalities of Short-Range Order (SRO) Calculation

For the analysis of the chemical ordering of a nanoalloy, we calculated the average short-range order (SRO) parameter, as defined by Cowley, for all the atoms in our nanoparticle. To examine potential variations of SRO with distance, the SRO parameter was calculated within the first four nearest-neighbour shells of each atom. This is usually done by defining concentric spherical shells around each atom of such radii that in an undistorted crystal lattice they would contain each atom's first to fourth nearest-neighbours, respectively. However, since our nanoparticles are highly distorted due to straining, this definition would not work well.

Therefore, we defined the shells differently. Knowing that our nanoparticle was *fcc*, we knew the coordination number, and, hence, the number of nearest-neighbours in each shell: namely, 12/6/12/8 nearest-neighbours for the first four shells, respectively. Also running simulations well below the nanoparticle melting points, as confirmed from OVITO visualisation, the crystalline nature of the relaxed nanoparticles was retained even within the highly distorted regions after straining. Then, we could define each nearest-neighbour shell as the space around each atom containing 12/6/12/8 neighbours, regardless of the actual distance between these atoms and the reference atom.

This definition of nearest-neighbour shells created an additional complication related to the surface atoms, which have fewer nearest-neighbours. These atoms could not be considered during scanning of the nanoparticle, as their coordination number was lower than that of bulk atoms and their nearest-neighbour shells would be miscalculated. Still, surface atoms needed to be considered during searching for nearest-neighbours, since they were located within nearest-neighbour shells of atoms in sub-surface positions. The way to circumvent this problem was by creating two files of atomic coordinates: one containing reference atoms (which excluded surface atoms), and one containing atoms to be examined for potential nearest-neighbour relationships (which included surface atoms).

---